\documentclass[epj,final]{svjour}
\usepackage{color}
\usepackage{graphics}
\usepackage{graphicx}
\def\lsim{\mathrel{\rlap{
\lower4pt\hbox{\hskip-3pt$\sim$}}
    \raise1pt\hbox{$<$}}}     
\def\gsim{\mathrel{\rlap{
\lower4pt\hbox{\hskip-3pt$\sim$}}
    \raise1pt\hbox{$>$}}}     
\def\scr#1{\mbox{\scriptsize #1}}

\begin{document}

\title{What can we learn from the directed flow 
in heavy-ion collisions at BES RHIC energies?}
\titlerunning{What can we learn from the directed flow}
\author{
Yu. B. Ivanov\inst{1,2}   \and 
A. A. Soldatov\inst{2}} 
\institute{                     
National Research Centre "Kurchatov Institute" (NRC "Kurchatov Institute"), 
Moscow 123182, Russia
\and
National Research Nuclear University "MEPhI"  (Moscow Engineering
Physics Institute), 
Moscow 115409, Russia
}
\date{Received: date / Revised version: date}
%
\abstract{
Analysis of directed flow ($v_1$) of protons,  antiprotons and pions in heavy-ion collisions
is performed   in the range of collision energies  $\sqrt{s_{NN}}$ = 2.7--39 GeV. 
Simulations have been done within a three-fluid model
employing a purely hadronic equation of state (EoS) and two versions of the EoS
with deconfinement transitions: 
a first-order phase transition and a smooth crossover transition.
The crossover EoS is unambiguously preferable for the description of 
the most part of experimental data in this energy range. 
The directed flow indicates that the crossover deconfinement transition
takes place in semicentral Au+Au collisions in a wide range of
collision energies  4 $\lsim\sqrt{s_{NN}}\lsim$ 30 GeV. 
The obtained results   
suggest that the deconfinement  EoS's in the quark-gluon sector should be 
stiffer at high baryon densities than those used in the calculation. 
The latter finding is in agreement with that discussed in astrophysics. 
}


\maketitle

\section{Introduction}

The directed flow \cite{Danielewicz:1985hn} is one of the key
observables in heavy ion collisions. Nowadays it is defined as  
the first  coefficient, $v_1$, in the Fourier expansion of a particle distribution, ${d^2 N}/{d y\;d\phi}$,  
in azimuthal angle $\phi$ with respect to the reaction plane \cite{Voloshin:1994mz,Voloshin:2008dg}
\begin{eqnarray}
 \label{vn-def}
\frac{d^2 N}{d y\;d\phi} = \frac{d N}{dy}
\left(1+ \sum_{n=1}^{\infty} 2\; v_n(y) \cos(n\phi)\right),
\end{eqnarray}
where $y$ is the longitudinal rapidity of a particle. 
The directed flow is mainly formed at an early
(compression) stage of the collisions and hence is sensitive to
early pressure gradients in the evolving nuclear matter \cite{So97,HWW99}. 
As the EoS is harder, stronger pressure is developed.
Thus the directed flow probes the stiffness of the nuclear EoS 
at the early stage of nuclear collisions \cite{RI06},
which is of prime interest for heavy-ion research.

In Refs. \cite{HS94,Ri95,Ri96}, 
a significant reduction of the directed flow 
in the first-order phase transition to the quark-gluon phase (QGP)
(the so-called "softest-point" effect)
was predicted, which results from decreasing
the pressure gradients in the
mixed phase as compared to those in pure hadronic and quark-gluon phases.
It was further predicted  \cite{CR99,Br00,St05} that the directed flow as a function of rapidity exhibits a wiggle near the midrapidity 
with a negative slope near the midrapidity, when the incident energy is in the range  
corresponding to onset of the first-order phase transition. 
Thus, the  wiggle near the midrapidity and the wiggle-like behavior 
of the excitation function of the midrapidity $v_1$ slope were put forward as 
a signature of the QGP phase transition. 
However, the QGP EoS is not a necessarily
prerequisite for occurrence of the midrapidity $v_1$ wiggle
\cite{SSV00}:  
A certain combination of space-momentum correlations may result
in a negative slope in the rapidity dependence of the directed flow
in high-energy nucleus-nucleus collisions.

The directed flow of
identified hadrons---protons, antiprotons, positive and negative pions---in Au+Au collisions was recently measured in the energy range $\sqrt{s_{NN}}$ =(7.7-39) GeV
by the STAR collaboration within the framework of the beam energy scan (BES) program 
at the BNL Relativistic Heavy Ion Collider (RHIC)\cite{STAR-14}.
These data  have been already discussed in Refs. \cite{SAP14,Konchakovski:2014gda,Ivanov:2014ioa}. 
The Frankfurt group \cite{SAP14} did not succeed to describe
the data and to obtain conclusive results. Within a
hybrid approach \cite{Petersen:2008dd}, the authors found that there is no sensitivity of 
the directed flow on the EoS and,
in particular, on the occurrence of a first-order phase transition.
One of the possible reasons of this result can be that   
the initial stage of the collision in all scenarios 
is described within the Ultrarelativistic Quantum Molecular
Dynamics (UrQMD) \cite{Bass98}  in the hybrid approach. 
However, 
this initial stage does not solely determine the final directed flow because 
the UrQMD results still differ from those obtained within 
the hybrid approach \cite{Petersen:2008dd}.

In Refs. \cite{Konchakovski:2014gda,Ivanov:2014ioa}
the new STAR data were analyzed within 
two complementary approaches: kinetic transport approaches
of the parton-hadron string dynamics (PHSD) \cite{CB09}
and its purely hadronic version (HSD) \cite{PhysRep}),
and a hydrodynamic approach of the relativistic three-fluid dynamics (3FD) \cite{IRT06,Iv13-alt1}.
In contrast to other observables, the
directed flow was found to be very sensitive to the accuracy settings of the numerical scheme. 
Accurate calculations require a very high memory and computation time. 
Due to this reason  in Refs. \cite{Konchakovski:2014gda,Ivanov:2014ioa}  we failed 
to perform calculations for energies above $\sqrt{s_{NN}}=$ 30
GeV within the 3FD model. 

In the present paper we extend the analysis performed in Refs. \cite{Konchakovski:2014gda,Ivanov:2014ioa} 
within the 3FD model. We succeeded to perform simulations at 
$\sqrt{s_{NN}}=$ 39 GeV, which allow us to strengthen conclusions on the relevance 
of used EoS's, in particular, on the stiffness of the EoS at high baryon densities 
in the QGP sector and outline the range of the crossover transition to the QGP in terms of 
collision energy.

\section{The 3FD model}
\label{sec:3FD}

The 3FD approximation is a minimal way to simulate
the early-stage nonequilibrium state of the colliding nuclei 
at high incident energies. 
The 3FD model~\cite{IRT06} describes a nuclear collision from the stage of 
the incident cold nuclei approaching each other, to the
final freeze-out stage. Contrary to the conventional one-fluid dynamics,
where a local instantaneous stopping of matter of the colliding nuclei
is assumed, the 3FD considers inter-penetrating counter-streaming flows of leading baryon-rich
matter, which gradually decelerate each other due to mutual friction.
The basic idea of a 3FD approximation to heavy-ion
collisions is that  a
generally nonequilibrium distribution of baryon-rich matter 
at each space-time point can be
represented as a sum of two distinct contributions initially
associated with constituent nucleons of the projectile and target
nuclei. In addition, newly produced particles, populating
predominantly the midrapidity region, are attributed to a third, so-called fireball
fluid that is governed by the net-baryon-free sector of the EoS.

At the final stage of the collision the p- and t-fluids
are either spatially separated or mutually stopped and unified,   
while the 
f-fluid, predominantly located in the midrapidity region,
keeps its identity and 
still overlaps with the baryon-rich fluids to a
lesser (at high energies) or greater (at lower energies) extent.
In particular, the friction between the baryon-rich and 
net-baryon-free fluids is the only source of dissipation at this final stage
because each separate fluid is described by equations of the ideal hydrodynamics 
with the right-hand sides containing friction with other fluids. 
All the observables in the 3FD model are composed of contributions
from all three fluids. 
Of course, relative fractions of these contributions depend on multiple factors, 
in particular, on the rapidity range. 
The freeze-out  is performed accordingly to the procedure described in Ref. \cite{IRT06} 
and in more detail in Refs. \cite{Russkikh:2006aa,Ivanov:2008zi}. 
This is a so-call modified Milekhin scheme that conserves the energy, momentum and baryon number. 
The freeze-out   criterion is based on a local    
energy density, $\varepsilon_{\scr{tot}}$,  defined as a sum of contributions 
from all (p-, t- and f-) fluids being present in a local space-time region. 
The freeze-out procedure starts if $\varepsilon_{\scr tot}<\varepsilon_{\scr{frz}}$.
The freeze-out energy density $\varepsilon_{\scr{frz}}$ = 0.4 GeV/fm$^3$ was chosen mostly
on the condition of the best reproduction of secondary
particle yields.

\begin{figure}[tbh]
\centerline{
\includegraphics[width=6.2cm]{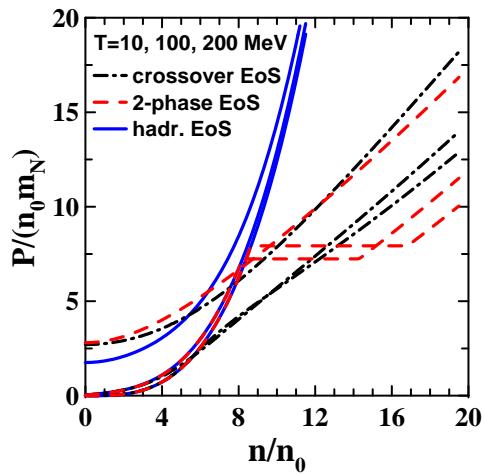}
}
 \caption{ 
Pressure scaled by the product of normal nuclear density ($n_0=$ 0.15 fm$^{-3}$) and 
nucleon mass ($m_N$) versus baryon density (scaled by $n_0$)
for three considered EoS's. Results are presented for three different
temperatures $T=$ 10, 100 and 200 MeV (bottom-up for corresponding curves).  
} 
\label{fig1.0}
\end{figure}

Different EoS's can be implemented in the 3FD model. 
A key point is that the 3FD model is able to treat a deconfinement transition at the
early {\em nonequilibrium} stage of the collision, when  
the directed flow is mainly formed. 
In this work we apply a purely hadronic EoS~\cite{GM79}, an EoS with a
crossover transition as constructed in Ref.~\cite{KRST06}
and an EoS with a first-order phase transition into the QGP \cite{KRST06}. 
These are illustrated in Fig. \ref{fig1.0}. 
Note that an onset of deconfinement in the 2-phase EoS takes place at 
rather high baryon densities, above $n\sim 8~n_0$. 
In EoS's compatible with constraints on the occurrence of the quark matter phase in massive neutron stars, 
the phase coexistence starts at about $4~n_0$ \cite{Klahn:2013kga}. 
As it will be seen below, this excessive softness of the deconfinement EoS's of Ref.~\cite{KRST06}
is an obstacle for proper reproduction of the directed flow at high collision energies.

%

\begin{figure}[thb]
\centerline{
\includegraphics[width=6.0cm]{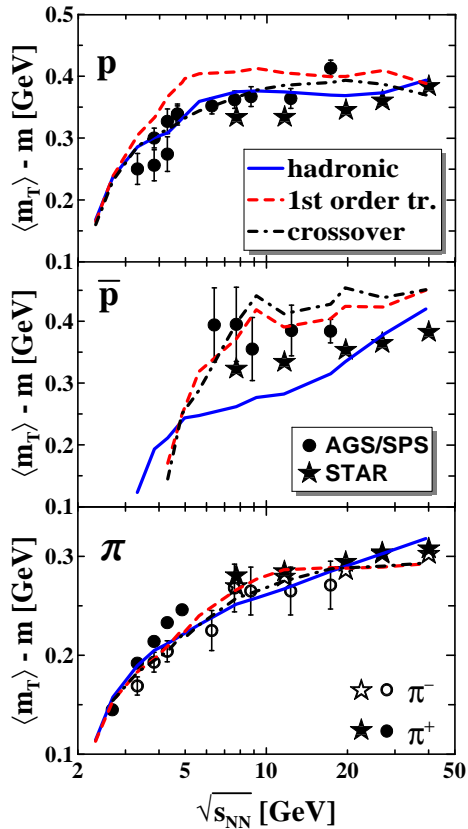}
}
 \caption{
Average transverse mass (minus particle mass) at midrapidity of protons ($p$), 
antiprotons ($\bar{p}$) and pions ($\pi$) 
from central ($b=$ 2 fm) Au+Au collisions 
as functions of the collision energy.   
Experimental data are taken from Refs. 
\cite{Ahle:1999uy,Klay:2001tf,Barrette:1999ry,Ahle:1999in,Ahle:1998jc}
for AGS, 
\cite{Afanasiev:2002mx,Alt:2007aa,Alt:2006dk,Anticic:2004yj}
for SPS and  \cite{Kumar:2014tca} for BES-RHIC energies.  
} 
\label{fig_mean-mt}
\end{figure}

In recent papers
\cite{Iv13-alt1,Iv13-alt2,Iv13-alt3,Ivanov:2012bh,Ivanov:2013cba,Ivanov:2013mxa,Iv14} 
a large variety of bulk observables has been analyzed
with these three EoS's:
the baryon stopping~\cite{Iv13-alt1,Ivanov:2012bh}, yields of different hadrons,
their rapidity and transverse momentum
distributions~\cite{Iv13-alt2,Iv13-alt3,Ivanov:2013cba}, 
the elliptic flow of various species \cite{Ivanov:2013mxa,Iv14}.
This analysis has been done in the
same range of incident energies as that in the present paper. 
Comparison with available data indicated a definite advantage of the  
deconfinement scenarios over the purely hadronic one 
especially at high collision energies.
As an example, excitation functions of average transverse masses 
($\langle m_T\rangle -m$) of  
protons, antiprotons and pions are presented in Fig. \ref{fig_mean-mt}.  
This quantity characterizes the transverse radial flow in the reaction and 
hence is relevant to the directed flow discussed here. 
In Fig. \ref{fig_mean-mt}
the 3FD results on $\langle m_T\rangle$ are  
confronted to old AGS/SPS data
\cite{Ahle:1999uy,Klay:2001tf,Barrette:1999ry,Ahle:1999in,Ahle:1998jc,Afanasiev:2002mx,Alt:2007aa,Alt:2006dk,Anticic:2004yj}
and also compared with new data from the STAR collaboration  \cite{Kumar:2014tca}. 
On the average, the crossover scenario still looks preferable, 
though these new STAR data are somewhat below those from SPS (for protons and antiprotons). 
It could seem that the crossover and hadronic scenarios reproduce (or fail to reproduce) 
the antiproton STAR data to the same extent. However, the slope of the crossover curve is 
definitely more adequate to the data. The first-order-transition scenario certainly fails for protons 
in the energy region where onset of deconfinement takes place in this EoS. 
The observed step-like behavior of the $\langle m_T\rangle$ excitation functions originates from 
peculiarities of the freeze-out in the 3FD model, 
as it is discussed in Ref. \cite{Iv13-alt3}  in detail. 
The physical pattern behind this freeze-out resembles the
process of expansion of compressed and heated classical
fluid into vacuum, mechanisms of which were studied both
experimentally and theoretically. The freeze-out is associated with evaporation from 
the surface of the expanding fluid and sometimes with the explosive 
transformation of the strongly superheated fluid to the gas.

The physical input
of the present 3FD calculations is described in detail in
Ref.~\cite{Iv13-alt1}.

\section{Results}

\begin{figure}[thb]
\includegraphics[width=0.48\textwidth]{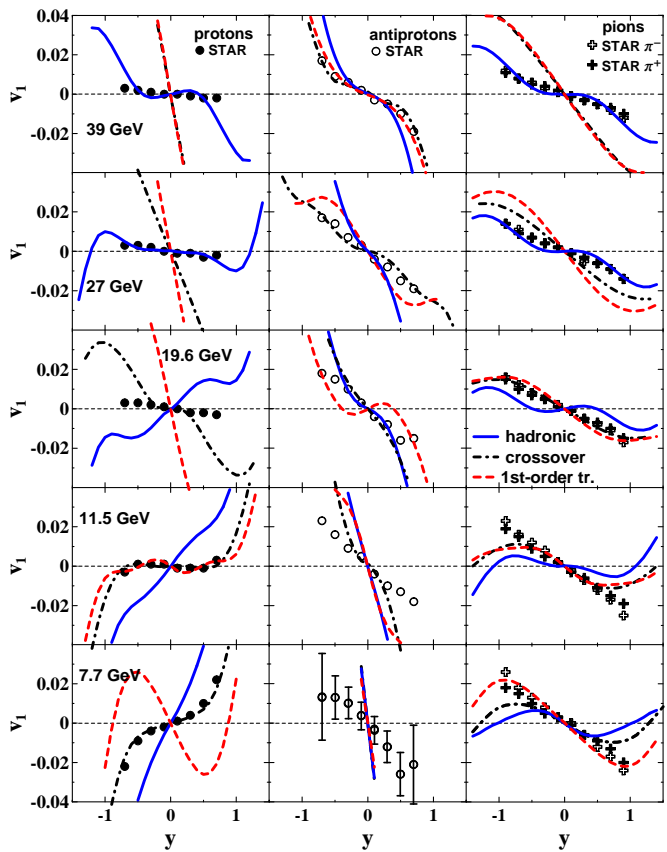}
\caption{ 
The directed flow $v_1(y)$ for protons,
  antiprotons and pions from mid-central ($b=$ 6 fm) Au+Au collisions at
  $\sqrt{s_{NN}}=$ 7.7--39 GeV
  calculated with different EoS's. Experimental data
  are from the STAR collaboration~\cite{STAR-14}.}
\label{fig:RHIC}
\end{figure}

The 3FD simulations were performed for mid-central Au+Au collisions, i.e. at impact parameter 
$b=$ 6 fm. 
In Ref. \cite{Abelev:2008ab} the values of the impact parameter corresponding 
to the experimental range of 10--40\% most central collisions 
were obtained using a Monte-Carlo Glauber calculation. 
The range of $b\approx$ 4.7--9.4 fm was found. This range  
manifested very weak collision-energy dependence. 
The mean value of this range is $\langle b\rangle\approx$ 7 fm, 
which is usually used for representative calculations for this 
centrality bin, see e.g.  PHSD/HSD calculations in Ref. \cite{Konchakovski:2014gda}. 
The Glauber calculation \cite{Abelev:2008ab} was based on 
a Woods-Saxon density profile 
with the diffuseness parameter $\approx$0.5 fm and resulted in 
  the total hadronic cross-section for Au+Au collisions 
$\sigma_{AA}\simeq$ 720 barns. 
Within the 3FD model the initial nuclei are represented
by sharp-edged spheres, i.e. with zero diffuseness.
This is done for stability of the incident nuclei before
collision \cite{IRT06}. Under this approximation 
the total hadronic Au+Au cross-section turns out to be 
$\sigma_{AA}^{\rm{3FD}}\simeq$ 580 barns. Therefore, 
we have to accordingly scale the mean value of the impact parameter as 
$\langle b^{\rm{3FD}} \rangle\approx \langle b\rangle (\sigma_{AA}^{\rm{3FD}}/\sigma_{AA})^{1/2}\approx $ 
6 fm. This justifies our choice of the representative $b$.  
Following the experimental conditions, the acceptance
$p_T <$ 2 GeV/c for transverse momentum ($p_T$) of the produced particles 
is applied to all considered hadrons. 
The directed flow $v_1(y)$ as a function of rapidity $y$ at BES-RHIC
bombarding energies is presented in Fig.~\ref{fig:RHIC} for
pions, protons and antiprotons.

As seen, the first-order-transition scenario gives results  for the proton $v_1$ which 
strongly differ from those in the crossover scenario at  $\sqrt{s_{NN}}=$ 7.7 and 19.6 GeV. 
This is in contrast to other bulk observables analyzed so far 
\cite{Iv13-alt1,Iv13-alt2,Iv13-alt3,Ivanov:2012bh,Ivanov:2013cba,Ivanov:2013mxa,Iv14}. 
At $\sqrt{s_{NN}}=$ 39 GeV the directed flow of all considered species 
practically coincidence within the first-order-transition and crossover scenarios. 
It means that the crossover transition to the QGP has been practically completed 
at $\sqrt{s_{NN}}=$ 39 GeV. 
It also suggests that the region 7.7 $\leq\sqrt{s_{NN}}\leq$  30 GeV, 
where the crossover EoS provides the best (although not perfect) 
overall reproduction of the STAR data, is the region of the crossover transition.

The crossover EoS is definitely the best in 
reproduction of the proton $v_1(y)$ at $\sqrt{s_{NN}}\leq$  20 GeV. 
However, surprisingly the hadronic scenario becomes preferable for the proton $v_1(y)$
at $\sqrt{s_{NN}}>$  20 GeV.  
A similar situation takes place in the PHSD/HSD transport approach.
Indeed, predictions of the HSD model   
(i.e. without the deconfinement transition) for the proton $v_1(y)$ become preferable
at $\sqrt{s_{NN}}>$  30 GeV \cite{Konchakovski:2014gda}, 
i.e. at somewhat higher energies than in the 3FD model. 
Moreover, the proton $v_1$ predicted by the UrQMD model, as cited in the experimental
paper \cite{STAR-14} and in the recent theoretical work \cite{SAP14}, 
better reproduces the proton $v_1(y)$ data at high collision energies 
than the PHSD and 3FD-deconfinement models do. 
Note that the UrQMD model is based on the hadronic dynamics. 
All these observations could be considered as an evidence of a 
problem in the QGP sector of a EoS. At the same time 
the  antiproton directed flow at $\sqrt{s_{NN}}>$  10 GeV
definitely indicates a preference of the crossover scenario 
within both the  PHSD/HSD and 3FD approaches.

This puzzle has a natural resolution within the 3FD model.  
The the QGP sector of the EoS's with deconfinement  \cite{KRST06} was fitted to the lattice QCD 
data at zero net-baryon density and just extrapolated to nonzero baryon densities. 
The protons mainly originate from baryon-rich fluids that are governed by the EoS at finite baryon densities.
The too strong proton antiflow within the crossover scenario at $\sqrt{s_{NN}}>$ 20 GeV 
is a sign of too soft QGP EoS at high baryon densities.   
In general, the antiflow or a weak flow indicates softness of an EoS 
\cite{RI06,HS94,Ri95,Ri96,CR99,Br00,St05,SSV00}. 
Predictions of the first-order-transition EoS, 
the QGP sector of which is constructed in the same way as that of the crossover one, 
fail even at lower collision energies,
when the QGP starts to dominate in the collision dynamics, i.e. at $\sqrt{s_{NN}}\gsim$ 15 GeV.   
This fact also supports the conjecture on a too soft QGP sector 
at high baryon densities in the used EoS's.

At the same time, 
the net-baryon-free (fireball) fluid is governed by the EoS at zero net-baryon density.  
This fluid is a main source of antiprotons 
(more than 80\% near midrapidity at $\sqrt{s_{NN}}>$ 20 GeV and $b=$ 6 fm), 
$v_1(y)$ of which is in good agreement with the data at
$\sqrt{s_{NN}}>$ 20 GeV within the crossover scenario 
and even in perfect agreement -- within the first-order-transition scenario at $\sqrt{s_{NN}}=$ 39 GeV. 
It is encouraging because at zero net-baryon density the QGP sector of the EoS's is fitted to the lattice QCD 
data and therefore can be trusted. 
The crossover scenario, as well as all other scenarios, definitely fails to reproduce 
the antiproton $v_1(y)$ data at 7.7 GeV. 
The reason is low multiplicity of produced antiprotons. 
The antiproton multiplicity in the mid-central ($b=$ 6 fm) Au+Au collision
at 7.7 Gev is $~1$ within the deconfinement scenarios and $~3$ within the hadronic scenario. 
Therefore, the hydrodynamical approach based on the grand canonical ensemble is certainly 
inapplicable to the antiprotons in this case. 
The grand canonical ensemble, with respect
to conservation laws, gives a satisfactory description of abundant particle production in
heavy ion collisions. However, when applying the statistical treatment to rare probes 
one needs to treat the conservation laws exactly, that is the canonical approach. 
The exact conservation
of quantum numbers is known to reduce the phase
space available for particle production due to additional constraints appearing through
requirements of local quantum number conservation. An example of applying the 
canonical approach to the strangeness production can be found in \cite{Hamieh:2000tk} 
an references therein.

The pions are produced from all fluids:  near midrapidity 
$\sim 60\%$ from the baryon-rich fluids and $\sim 40\%$ from the net-baryon-free one at $\sqrt{s_{NN}}>$ 20 GeV. 
Hence, the disagreement of the pion $v_1$ with data, resulting from  redundant softness of the 
QGP EoS at high baryon densities, is moderate at $\sqrt{s_{NN}}>$ 20 GeV. 
In general, the pion $v_1$ is less sensitive to the EoS as compared to the 
proton and antiproton ones. 
As seen from Fig.~\ref{fig:RHIC}, the deconfinement scenarios are definitely 
preferable for the pion $v_1(y)$ at $\sqrt{s_{NN}}<$ 20 GeV. 
Though, the hadronic-scenario results are not too far from the experimental data. 
At $\sqrt{s_{NN}}=$ 39 GeV the hadronic scenario gives even the best description 
of the pion data because of a higher stiffness of the hadronic EoS 
at high baryon densities, as compared with that in the considered versions of the QGP EoS.  

Thus, all the analyzed data testify in favor of a harder QGP EoS at high baryon densities 
than those used in the simulations, i.e. the desired QGP EoS should be closer 
to the used hadronic EoS at the same baryon densities (see Fig. \ref{fig1.0}).  
At the same time, a moderate softenning 
of the QGP EoS at moderately high baryon densities 
is agreement with data at 
7.7 $\lsim\sqrt{s_{NN}}\lsim$ 20 GeV.

Here it is appropriate to mention a discussion on the QGP EoS in astrophysics. 
In Ref. \cite{Alford:2004pf} it was demonstrated that the QGP EoS can be 
almost indistinguishable from the hadronic EoS at high baryon densities relevant to neutron stars. 
In particular, this gives a possibility to explain hybrid stars with masses up to about 2 solar masses ($M_\odot$),
in such a way that ``hybrid stars masquerade as neutron stars'' \cite{Alford:2004pf}. 
The discussion of such a possibility has been revived after 
measurements on two binary pulsars PSR J1614-2230 \cite{Demorest:2010bx} 
and PSR J0348+0432 \cite{Antoniadis:2013pzd} 
resulted in the pulsar masses of (1.97$\pm$0.04)$M_\odot$ and (2.01$\pm$0.04)$M_\odot$, respectively. 
The obtained results on 
the directed flow give us another indication of a required hardening of the QGP EoS 
at high baryon densities.


\begin{figure}[thb]
\centerline{
\includegraphics[width=0.35\textwidth]{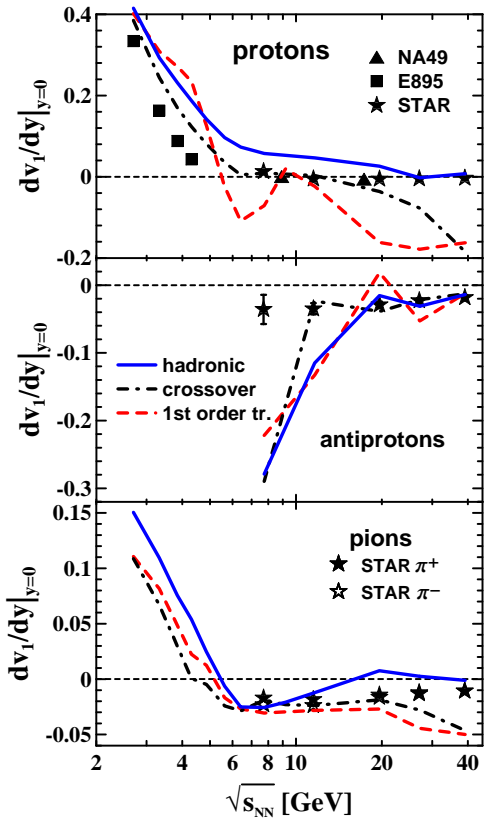}
}
\caption{ 
The beam energy dependence of the directed
  flow slope at midrapidity for protons, antiprotons and pions 
  from mid-central ($b=$ 6 fm) Au+Au collisions calculated
  with different EoS's.
   The experimental data are from Refs. \cite{STAR-14,NA49,E895}.}
\label{fig:slope}
\end{figure}

The slope of the directed flow at the midrapidity is often used to quantify variation 
of the directed flow with collision energy. 
The excitation functions  for the slopes of the $v_1$ distributions
at  midrapidity are summarized in Fig.~\ref{fig:slope}, where 
earlier experimental results from the AGS \cite{E895}  and SPS \cite{NA49}
are also presented. As noted
above, the best reproduction of the data is achieved with the 
crossover EoS. The proton $d v_1/dy$ within the first-order-transition scenario 
exhibits a wiggle earlier predicted in Refs. \cite{Ri95,Ri96,Br00,St05}.
The first-order-transition results demonstrate the worst 
agreement with the proton and antiproton data on $d v_1/dy$. 
The discrepancies between experiment and the 3FD 
predictions are smaller for the purely hadronic EoS, however, 
the agreement 
for the crossover EoS is definitely better
though it is far from being perfect. 
As seen from Fig.~\ref{fig:slope}, the rising transparency of colliding nuclei 
makes the proton  $v_1$ slope to be close to zero even for comparatively hard hadronic EoS.

The above discussed problems 
of the crossover scenario 
reveal themselves also in the $d v_1/dy$ plot. 
At high energies ($\sqrt{s_{NN}}>$ 20 GeV), the slopes also    
indicate that the used deconfinement  EoS's in the quark-gluon sector 
at zero baryon chemical potential are quite 
suitable for reproduction of the antiproton $d v_1/dy$ while those 
at high baryon densities (proton slope)
should be stiffer  in order to achieve better description of proton $d v_1/dy$.
A combined effect of this excessive softness of the QGP EoS and  the 
reducing baryon stopping results in more and more negative proton slopes 
at high collision energies. This is in line with the mechanism discussed in 
Ref. \cite{SSV00}. The pion flow partially follows the proton pattern, 
as discussed above. Therefore, the pion $v_1$ slope also becomes more negative 
with energy rise. 
As for the poor reproduction of the proton $v_1$ slope  at low energies ($\sqrt{s_{NN}}<$ 5 GeV), it is still 
questionable because the same data but in terms of the transverse in-plane 
momentum, $\langle P_x\rangle$, are almost perfectly 
reproduced by the crossover scenario \cite{Ivanov:2014ioa}. 
It is difficult to indicate the beginning of the crossover transition 
because the crossover results become preferable beginning with relatively 
low collision energies ($\sqrt{s_{NN}}>$ 3 GeV). However, 
the beginning of the crossover transition can be 
approximately pointed out as $\sqrt{s_{NN}}\simeq$ 4 GeV.

Of course, the 3FD model does not include all factors determining the directed flow. 
Initial-state fluctuations, which in particular make the directed flow to be 
nonzero even at midrapidity, are out of the scope of the 3FD approach. 
Apparently, these fluctuations can essentially affect the directed flow at 
high collision energies, when the experimental flow itself is very weak. 
Another point is so-called afterburner, i.e. the kinetic evolution after the 
the hydrodynamical freeze-out. This stage is absent in the conventional version 
of the 3FD. Recently  Iu. Karpenko constructed an event generator based on the output 
of the 3FD model. Thus constructed particle ensemble can be further evolved within the 
UrQMD model. Preliminary results \cite{Karpenko2015} show that such kind of the afterburner mainly 
affects the pion $v_1$ at peripheral rapidities. At $\sqrt{s_{NN}}<$ 5 GeV, 
the midrapidity region of the pion $v_1$ is also affected.

\section{Conclusions}
\label{sec:conclusions}

In conclusion, 
the crossover EoS is unambiguously preferable for the  
the most part of experimental data in the considered energy range,  
though this  description is not perfect. 
Based on the crossover EoS of Ref. \cite{KRST06}, 
the directed flow in semicentral Au+Au collisions
indicates that the crossover deconfinement transition
takes place  in 
the wide range incident energies  4 $\lsim\sqrt{s_{NN}}\lsim$ 30 GeV.
In part, this wide range could be a consequence of that the crossover transition
constructed in Ref. \cite{KRST06} is very smooth.
In this respect, this version of the crossover EoS certainly contradicts
results of the lattice QCD calculations, where a fast crossover,
at least at zero chemical potential, was found \cite{Aoki:2006we}.

At highest computed energies of $\sqrt{s_{NN}}>$ 20 GeV, 
the obtained results   
indicate that the deconfinement  EoS's in the QGP sector should be 
stiffer at high baryon densities than those used in the calculation, 
i.e. more similar to the purely hadronic EoS. This observation is in 
agreement with that discussed in astrophysics. 

In view of this similarity of the QGP and hadronic EoS's at high baryon densities 
a question arises if we can use the directed flow to detect the deconfinement transition. 
Qualitatively, a wiggle in the excitation function of the net-proton $v_1$ slope 
could be a signal of the QGP formation.  
This signal is unavoidable because of the EoS softening in the transition range from hadronic phase to the QGP. 
In fact, such a weak wiggle is observed in the STAR data \cite{STAR-14}.
However, this signal is still ambiguous because the corresponding antiflow can  
be of pure hadronic nature, as it was pointed out in Ref. \cite{SSV00}. 
Therefore, we are left with a routine way of comparing predictions of various EoS's 
(with and without deconfinement) with the v1 experimental data. This will hopefully 
give us a reliable evidence on the QGP formation.

Fruitful discussions with D. Blaschke, 
W. Cassing, V. P. Konchakovski,  V. D. Toneev, and D.N. Voskresensky 
are gratefully acknowledged. 
We are grateful to A.S. Khvorostukhin, V.V. Skokov, and  V.D. Toneev  for providing 
us with the tabulated first-order-phase-transition and crossover EoS's. 
The calculations were performed at the computer cluster of GSI (Darmstadt). 
This work was partially supported by grant NS-932.2014.2.


\end{document}